\documentclass[12pt, a4paper]{JHEP}

\usepackage{epsfig}
\usepackage{latexsym}

\usepackage{psfrag}

\def\b{\beta}

\def\f{\phi}               
\def\g{\gamma}
\def\h{\eta}

\def\x{\xi}

\def\ch{{\cal H}}

\def\be{\begin{equation}}       \def\eq{\begin{equation}}
\def\ee{\end{equation}}         \def\eqe{\end{equation}}

\def\bea{\begin{eqnarray}}      \def\eqa{\begin{eqnarray}}
\def\ena{\end{eqnarray}}        \def\eea{\end{eqnarray}}
                                \def\eqae{\end{eqnarray}}

\newcommand{\hp}{{\widehat\Phi}}
\newcommand{\hq}{{\widehat Q_B}}
\newcommand{\he}{{\widehat\eta_0}}

\newcommand{\lll}{\big\langle\big\langle}
\newcommand{\rrr}{\big\rangle\big\rangle}
\newcommand{\lllb}{\Bigl\langle\Bigl\langle}
\newcommand{\rrrb}{\Bigr\rangle\Bigr\rangle}

\newcommand{\STRUT}{\rule{0in}{2ex}}

\title{Level-four approximation to the tachyon potential in superstring field
theory}

\author{Pieter-Jan De Smet and Joris Raeymaekers\\
Instituut voor theoretische fysica, Katholieke Universiteit Leuven,\\
Celestijnenlaan 200D, B-3001 Leuven, Belgium\\
E-mail: \email{Joris.Raeymaekers@fys.kuleuven.ac.be}, 
\email{Pieter-Jan.DeSmet@fys.kuleuven.ac.be}}

\abstract{We compute the tachyon potential to level $4$ in NS
superstring field theory.  We obtain $89 \%$ of the conjectured vacuum
energy.}

\received{April 6, 2000}
\accepted{May 29, 2000}
\keywords{D-branes, Superstring Vacua}
\preprint{ KUL-TF-2000/10 \\
  {\tt hep-th/0003220}}

\begin{document}

\section{Introduction}

It has been conjectured by Sen that at the stationary point of the
tachyon potential for the D-brane-anti-D-brane pair or for the non-BPS
D-brane of superstring theories, the negative energy density precisely
cancels the brane tensions~\cite{9805019, 9805170, 9911116}.

For the D-brane of bosonic string theory, this conjecture has been
verified by Sen and Zwiebach~\cite{9912249} starting from Witten's
open string field theory~\cite{WITTENBSFT}. Using a level truncation
method which was proposed by Kostelecky and Samuel~\cite{KS} and
including fields up to level 4, they found a contribution of $ 99 \%$
of the expected value.  Subsequently Taylor and Moeller continued the
calculation to level 10 fields and verified the conjecture to $99,9
\%$ \cite{moeller}.

In the superstring case the first calculation was done by
Berkovits~\cite{0001084} using his Wess-Zumino-Witten like proposal
for the string field theory action~\cite{9503099, 9912121}.  The pure
tachyon contribution was found to amount to $60 \%$ of the conjectured
value.  This computation was then continued to higher levels by the
combined force of Berkovits, Sen and Zwiebach~\cite{BSZ}.  They
included fields up to level $3/2$ and got $85 \%$ of the expected
answer.

In this paper we perform a further check on the conjecture: we
continue the calculation retaining fields up to level 2 and get $89 \%
$ of the conjectured value.

\section{Berkovits' superstring field theory}

Using the embedding of the $N=1$ superstring into a critical $N=2$
theory found in~\cite{vafa}, Berkovits proposed a superstring field
theory based on a Wess-Zumino-Witten type
action~\cite{9503099,9912121}. With slight modifications, this action
can be used to describe the NS-sector excitations of a non-BPS
brane~\cite{0001084} (the modification needed to include the R-sector
fields is as yet unknown).  In this section we briefly review the
action and some of its properties. This section summarizes parts
of~\cite{BSZ} relevant to the problem at hand with some additional
comments.

A string field describing an NS excitation on a non-BPS D-brane can be
represented by a vertex operator of the form
\begin{equation}
\widehat\Phi = \Phi\otimes(I \mbox{   or   } \sigma_1)
\end{equation}
where $\Phi$ is an operator in the conformal field theory of the NS
superstring with the superghost system bosonized as $\b = \partial \x
e^{-\f}$ and $\g = \h e^\f$~\cite{FMS}.\footnote{We adopt the
convention that $e ^{q \f}$ is a fermion for odd values of $q$, i.e.\
it anticommutes not only with $e^{q' \f}$ for odd $q'$ but also with
the other fermionic fields in the theory.}  $\Phi$ is restricted to
have ghost number~0, picture number~0 and to live in the `large'
Hilbert space which includes the $\x$ zero mode.  The string field
$\widehat\Phi$ should include states of both the GSO-projected and
GSO-unprojected sectors.  Fields in the GSO-unprojected sector are
tensored with $\sigma_1$ and the fields in the GSO-projected sector
are tensored with $I$.

One further defines $\widehat \eta_0 = \eta_0 \otimes \sigma_3$ where
$\eta_0$ is the zero-mode of the $\eta$-field, and $\widehat Q_B = Q_B
\otimes \sigma_3 $ where $Q_B$ is the BRST-charge
\be 
\label{eb4}
Q_B = \oint dz j_B(z) = \oint dz \Bigl\{ c \bigl( T_m + T_{\xi\eta} +
T_\phi) + c \partial c b +\eta \,e^\phi \, G_m - \eta\partial \eta
e^{2\phi} b \Bigr\}\, ,
\ee
and
\be 
\label{eb5}
T_{\xi\eta}=\partial\xi\,\eta, \quad T_\phi=-{1\over 2} \partial\phi
\partial \phi -\partial^2\phi \, ,
\ee
$T_m$ is the matter stress tensor and $G_m$ is the matter
supercurrent.  The string field action for the non-BPS D-brane then
takes the following form:
\begin{equation}
S=\frac{1}{4 g^2} \lllb (e^{-\hp} \hq e^{\hp})(e^{-\hp}\he e^\hp) -
\int_0^1 dt (e^{-t\hp}\partial_t e^{t\hp})\{ (e^{-t\hp}\hq e^{t\hp}),
(e^{-t\hp}\he e^{t\hp})\}\rrrb .
\label{action}
\end{equation}
With the double brackets we mean the following:
\begin{equation}
\lll \widehat A_1\cdots \widehat A_n \rrr = \left(\mbox{Trace of
matrices}\right) 
 \Bigl\langle f^{(n)}_1 \circ A_1(0)\cdots
f^{(n)}_n\circ A_n(0)\Bigr\rangle\,. 
\end{equation}
The $f_k^{(n)}$ entering in the correlator on the right hand side
denote some appropriate conformal transformations~\cite{leclair}.
They are defined by
\begin{equation}
f^{(n)}_k(z) = e^{2\pi i (k-1)\over n} \left({1+iz\over 1-iz}
\right)^{2/n}\,,\qquad  \hbox{for} \ n\geq 1\,,   
\end{equation}
they map the unit circle to wedge-formed pieces of the complex-plane.
\pagebreak[3]

One can show that the action (\ref{action}) is invariant under the
gauge transformation 
\be 
\label{egtrs} 
\delta e^{\widehat\Phi} =
\left(\widehat Q_B \widehat\Omega\right) e^{\widehat\Phi} +
e^{\widehat\Phi}\left(\widehat\eta_0 \widehat\Omega'\right) , 
\ee 
where $\widehat\Omega$ and $\widehat\Omega'$ are independent gauge
parameters.  This gauge invariance can be fixed\footnote{This is a
reachable gauge choice (if $L_0 \neq 0$) but we have not been able to
prove that it fixes the gauge freedom completely.}  by imposing
\begin{equation}
\label{gauge fixing}
b_0|\mbox{State}\rangle = 0 \mbox{   and   } \xi_0|\mbox{State}\rangle = 0\,.
\end{equation}

In the calculation of the tachyon potential, we can restrict the
string field to lie in a subspace $ \ch_1$ formed by acting only with
modes of the stress-energy tensor, the supercurrent and the ghost
fields $b$, $c$, $\eta$, $\x$, $\phi$, since the other excitations can be
consistently put to zero.  Furthermore, when restricted to fields
lying in $ \ch_1$ the action has a $Z_2$ twist invariance under which
the fields in the GSO-odd sector carry charge $(-)^{h+1}$ and the
fields in the GSO-even sector carry charge $(-)^{h+1/2}$, $h$ is the
conformal weight.  In the computation of the tachyon potential we can
therefore further restrict ourselves to the twist even
fields.\footnote{This restriction projects out the only state with
$L_0 = 0$, namely $\xi\partial\xi\ c\partial c\ e^{-2\phi}$.}

The non-polynomial action~(\ref{action}) should be formally expanded
in the string field $\widehat\Phi$, and each term should be
accompanied by the appropriate conformal transformations. However,
because we will only compute the interactions between a finite number
of fields, it is easy to see that one does not need all the terms in
the action. The conformal field theory correlators in the action
(\ref{action}) are nonvanishing only if the total $(b,\ c )$ number is
$3$, the $(\h,\ \x)$ number is 1 and the total $\f$-charge is $-2$. In
the following we will need only the terms in the action involving up
to $6$ string fields.

Making use of the cyclicity properties derived in the appendix
of~\cite{BSZ}, the action to this order can be written as
\begin{eqnarray*}
S &=& \frac{1}{2g^2} \Biggl\langle\Biggl\langle\frac{1}{2} (\hq \hp)(\he \hp) +
\frac{1}{3} (\hq\hp) \hp (\he\hp) + \frac{1}{12} (\hq\hp) \Bigl( \hp^2
(\he\hp) - \hp(\he\hp)\hp \Bigr)+
\\ && 
	\hphantom{ \frac{1}{2g^2} \lllb }\!
+\frac{1}{60} (\hq\hp) \Bigl( \hp^3 (\he\hp) 
- 3\hp^2(\he\hp)\hp \Bigr) +
\\&& 
	\hphantom{ \frac{1}{2g^2} \lllb }\!
+ \frac{1}{360}(\hq\hp) \Bigl( \hp^4 (\he\hp) - 4\hp^3(\he\hp)\hp +
3\hp^2(\he\hp)\hp^2 \Bigr) \Biggr\rangle\Biggr\rangle\,.
\label{truncaction}
\end{eqnarray*}

\section{The fields up to level 2}

Taking all this together we get the list of contributing
fields up to level 2 (table~\ref{t1}). The level of a field is just the conformal
weight shifted by 1/2. In this way the tachyon is a level 0 field. We
use the notation $|q\rangle$ for the state corresponding with the
operator $e^{q \phi}$.

The level $0$ and level $2$ fields should be tensored with $\sigma_1$ 
and the level $3/2$ fields with~$I$.

We list the conformal transformations of the fields in
appendix~\ref{transforms}.

\begin{table}[h]
\caption{The contributing states up to level two and their
corresponding vertex operators.\label{t1}}
\begin{center}
\begin{tabular}{|l|c|c|}\hline
\mbox {Level}
     & \mbox {state}& \mbox {vertex\ operator} \\
\hline     
0&$\xi_0 c_1|-1\rangle$&$ T = \xi c e^{-\phi}$\\ 
\hline
3/2\STRUT &$ 2 c_1 c_{-1} \xi_0\xi_{-1} |-2\rangle$
            &$A = c\partial^2 c\xi\partial\xi\ e^{-2\phi}$\\
    &$ \xi_0 \eta_{-1}  |0\rangle$&$E =  \xi\eta$\\
    &$  \xi_0 c_1 G^m_{-3/2}  |-1\rangle$&$F = \xi c G^m\ e^{-\phi}$\\ 
\hline
2\STRUT &$ \xi_0 c_1 \left[(\phi_{-1})^2-\phi_{-2}\right]|-1\rangle$
  &$ K = \xi c \ \partial^2\left(e^{-\phi}\right)$\\
  &$ \xi_0 c_1 \phi_{-2} |-1\rangle$&$ L = \xi c\ \partial^2\phi\ 
e^{-\phi}$\\
 &$\xi_0 c_1 L^m_{-2}  |-1\rangle$&$ M =\xi c T^m\ e^{-\phi}$\\
 &$2 \xi_0 c_{-1}   |-1\rangle$& $N = \xi \partial^2 c\ e^{-\phi}$\\
 &$\xi_0 \xi_{-1}\eta_{-1}c_1  |-1\rangle$&$ P = \xi\partial\xi\eta c\ 
e^{-\phi}$\\\hline
\end{tabular}
\end{center}
\end{table}

\section{The tachyon potential}

We have calculated the tachyon potential involving fields up to level
2, including only the terms up to level 4 (the level of a term in the
potential is defined to be the sum of the levels of the fields
entering into it).  We have performed the actual computation in the
following way.  The conformal tranformations of the fields were
calculated by hand.  The computation of all the correlation functions
between these transformed fields was done with the help of
Mathematica: we have written a program to compute the necessary CFT
correlation functions. We have performed an extra check by calculating
some of the correlators on the upper half plane instead of the disc.
Denoting
\begin{equation}
\widehat\Phi = t \widehat T + a \widehat A +e \widehat E + f \widehat
F+ k \widehat K + l \widehat L +m \widehat M+n \widehat N + p \widehat P
\end{equation}
we give the result with coefficients evaluated numerically up to 6
significant digits (subscripts refer to the level)
\begin{eqnarray*}
V(\widehat\Phi )&=& -S(\widehat\Phi)
\\*
&=& V_0 + V_{3/2} + V_2 + V_3 + V_{7/2} + V_4
\\[4pt]
V(\widehat\Phi )_0 
&=& 2 \pi^2 M (-0.25 t^2 + 0.5  t^4) 
\\[4pt] 
V(\widehat\Phi )_{3/2} &=&  2 \pi^2 M (-   a {t^2}-0.25 e {t^2}-  
0.518729 e {t^4}) 
\\[4pt]
V(\widehat\Phi )_2 &=&2 \pi^2M \Bigl( 0.333333 k {t^3} + 1.83333 l {t^3} -
 3.75 m {t^3} +2.83333 n {t^3} +0.25 p {t^3}\Bigr)
\\[4pt]
V(\widehat\Phi )_3 &=&2 \pi^2M \Bigl( 2 a e+5 {f^2}+ 4.96405 a e
{t^2}-0.66544 {e^2} {t^2}+
\\*& & 
	\hphantom{2 \pi^2 M (}\!
+ 5.47589 e f {t^2}+5.82107 {f^2} {t^2}+0.277778 {e^2} {t^4}\Bigr)
\\[4pt]
V(\widehat\Phi )_{7/2} &=&2 \pi^2 M \Bigl(-3.03704 a k t -7.11111 a l
t+2.77778 a m t -1.62963 a n t -
\\*& & 
	\hphantom{2 \pi^2 M (}\!
-1.55556 a p t+0.12963 e k t  -0.296296 e l t+0.694444 e m t-
\\*& & 
	\hphantom{2 \pi^2 M (}\!
-1.2963 e n t +0.944444 e p t-11.8519 f l t-8.88889 f m t -
\\*& &
	\hphantom{2 \pi^2 M (}\!
-2.96296 f p t  - 2.87299 e k {t^3} - 1.94348 e l {t^3} +4.35732 e m {t^3}-
\\*& &
	\hphantom{2 \pi^2 M (}\!
 -4.77364 e n {t^3} +0.605194 e p {t^3}\Bigr) 
\\ [4pt]
V(\widehat\Phi )_4 &=& 2 \pi^2 M \Bigl( 3 {k^2} -3 k l +1.5 {l^2}+5.625
{m^2}-3 {n^2} -0.75 {p^2}+10.3958 k^2 {t^2}+
\\*& &
	\hphantom{2 \pi^2 M (}\!
+0.791667 k l {t^2} -1.875 k m {t^2} +5.54167 k n
{t^2} - 1.4375 k p {t^2}+
\\*& &
	\hphantom{2 \pi^2 M (}\!
+6.70833 {l^2} {t^2}-10.3125 l m {t^2}
+ 11.9167 l n  {t^2} -0.875 l p {t^2}+
\\*& & 
	\hphantom{2 \pi^2 M (}\!
+14.7656 {m^2} {t^2}-15.9375 m n {t^2} - 1.40625 m p {t^2}+
\\*& & 
	\hphantom{2 \pi^2 M (}\!
+5.83333  n^2 {t^2} -0.5 n p {t^2} -1.5 {p^2} {t^2}\Bigr)\,.
\end{eqnarray*}
Our results for $V_0$, $V_{3/2}$ and $V_3$ agree completely
with~\cite{BSZ}\footnote{Note that our field $F$ is defined with a
different sign from the one in~\cite{BSZ}}.

This potential has extrema at $(\pm
t_0,a_0,e_0,f_0,\pm k_0,\pm l_0,\pm m_0,\pm n_0,\pm p_0)$ with 
$$
\begin{array}{rclcrclcrcl}
t_0&=& 0.602101\\
a_0&=& 0.0521934\,,&\qquad& e_0 &=& 0.0430366\,,&\qquad& f_0 &=& -0.0138164\,, \\
k_0 &=& -0.01019\,,&\qquad& l_0 &=&  -0.0450433\,,&\qquad& m_0 &=& 0.0322127\,,\\ 
n_0 &=& 0.0473113\,,&\qquad& p_0 &=& 0.021291\,.
\end{array}
$$
At these extrema
\begin{equation}
V = -0.891287 \, M\,,
\end{equation}
so we see that at the level $4$ we get $89 \%$ of the conjectured
 value $V= - M$ ($M$ is the D-brane mass).  In the potential computed to
 level $4$ all the fields but $t$ appear only quadratically.  They can
 be integrated out very easily to give the effective potential $V(t)$
 see figure~\ref{f1}.

\FIGURE[t]{\small\begin{picture}(300,195)
\put(135,190){$V(t)/M$}
\put(295,80){$t$}
\put(0,0){\epsfig{file=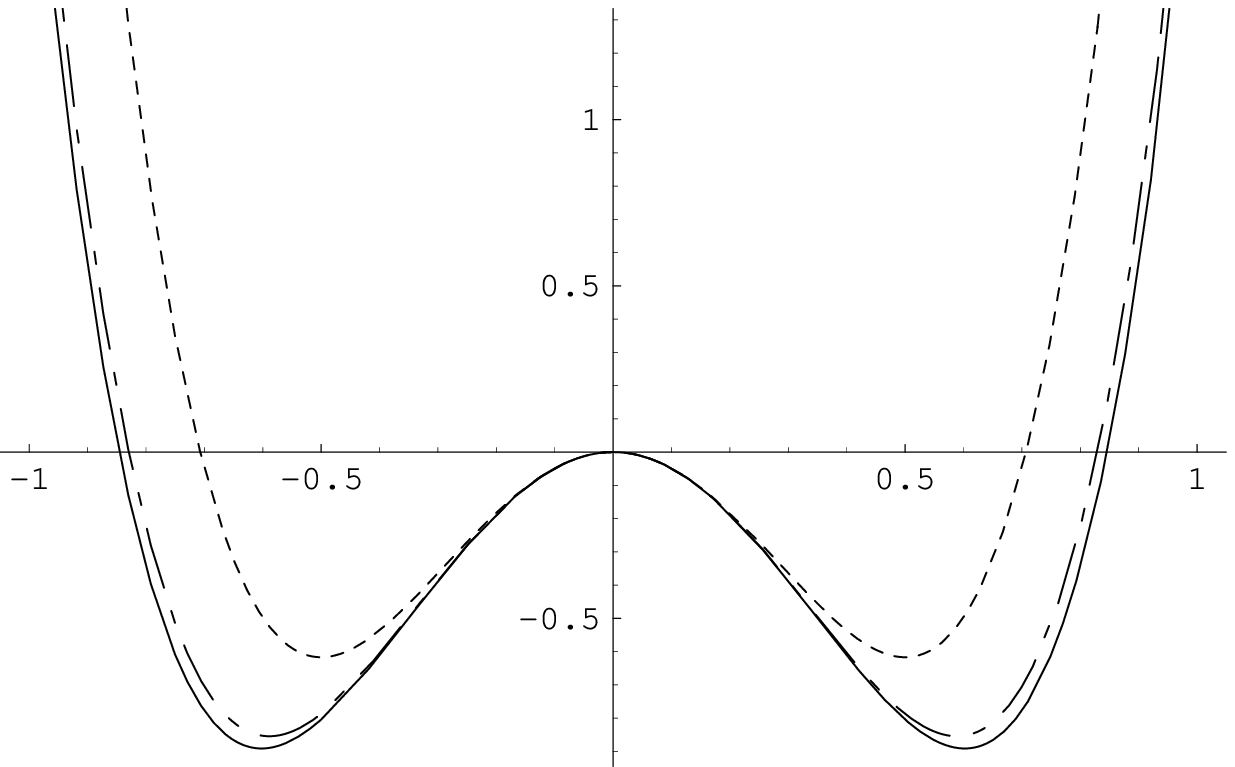, width=300pt}}
	      \end{picture}%
\caption{The tachyon potential $V(t)/M$ at level 0 (dotted line), 
level 3 (dashed line)  and level 4 (full line).\label{f1}}}

\section{Conclusions and outlook}

The last few months have seen an increased confidence in string field
theory as a calculational tool for doing off-shell string
calculations.  In this letter, we have used Berkovits' superstring
field theory to calculate the tachyon potential up to level four. Our
result shows a further convergence towards the total vacuum energy
conjectured by Sen, albeit less rapid than the contributions of the
previous levels. Therefore it would be interesting to persue the
calculation to higher levels.  At present, not much is
known about the general convergence properties of level-truncation
calculations in superstring theory. It would be nice to have a deeper
understanding of this.  Another interesting problem would be to study
the interactions of the massless vector using Berkovits' action and
compare with the different proposals for the non-abelian
Dirac-Born-Infeld action~\cite{0001201}.
   
\acknowledgments{This work was supported in part by the European
Commission TMR project\linebreak ERBFMRXCT96-0045.  We would like to thank
N.~Berkovits and W.~Taylor for encouragement.  We have benefitted from
discussions with B.~Craps, F.~Roose, J.~Troost and W.~Troost, and from
correspondence with Amer Iqbal and Asad Naqvi.  P.J.D.S. is aspirant
FWO-Vlaanderen.}

\appendix
\section{The conformal transformations of the
fields}\label{transforms}

We now list the conformal transformations of the fields used in the calculation
of the tachyon potential. To shorten the notation we denote $w = f(z)$.
\begin{eqnarray*}
f \circ T(z) &=& (f'(z))^{-1/2} T(w)
\\
f \circ A(z) &=& f'(z) A(w) -
      \frac{f''(z)}{f'(z)}c\partial c\ \xi\partial\xi\ e^{-2 \phi}(w)
\\
f \circ E(z) &=& f'(z) E(w) -\frac{f''(z)}{2 f'(z)}
\\
f \circ F(z) &=& f'(z) F(w)
\\
f \circ K(z) &=& (f'(z))^{3/2} K(w) + 
2 \frac{f''(z)}{f'(z)} (f'(z))^{1/2}\xi c
\partial\left( e^{-\phi}\right)(w)+
\\*& &   
+ \left[ \frac{1}{2}\frac{f'''}{f'}-\frac{1}{4}
\left(\frac{f''}{f'}\right)^2 \right] (f'(z))^{-1/2} \xi c
e^{-\phi}(w)
\\
f \circ L(z) &=& (f'(z))^{3/2} L(w) +  \frac{f''(z)}{f'(z)}(f'(z))^{1/2} 
 \xi c \partial\phi\ e^{-\phi}(w)+
\\*& & 
+ \left[ \frac{3}{4}\left(\frac{f''}{f'}\right)^2-
\frac{2}{3}\frac{f'''}{f'}\right] (f'(z))^{-1/2} 
\xi c e^{-\phi}(w)
\\
f \circ M(z) &=& (f'(z))^{3/2} M(w) + \frac{15}{12} \left[\frac{f'''}{
f'} - \frac{3}{2}\left( \frac{f''}{f'}\right)^2\right](f'(z))^{-1/2}
\xi c e^{-\phi}(w)
\\
f \circ N(z) &=& (f'(z))^{3/2} N(w) -
 \frac{f''(z)}{f'(z)}(f'(z))^{1/2} \xi \partial c\ e^{-\phi}(w)+
\\*& & 
+ \left[2 \left(\frac{f''}{f'}\right)^2- \frac{f'''}{f'}\right]
(f'(z))^{-1/2} \xi c\ e^{-\phi}(w)
\\
f \circ P(z) &=& (f'(z))^{3/2} P(w) +
\frac{1}{2}\frac{f''(z)}{f'(z)}(f'(z))^{1/2} \partial\xi\ c
e^{-\phi}(w)+
\\*& & 
+ \left[ \frac{1}{4} \left(\frac{f''}{f'}\right)^2-
\frac{1}{6}\frac{f'''}{f'}\right] (f'(z))^{-1/2} \xi c e^{-\phi}(w)
\end{eqnarray*}

\end{document}